\documentstyle[twoside,fleqn,espcrc2dsa2,epsf]{article}

\newcommand{\AmS}{{\protect\the\textfont2
  A\kern-.1667em\lower.5ex\hbox{M}\kern-.125emS}}

\def\eV{eV$/c^2$}
\def\cf{$^{252}$Cf}
\def\co{$^{60}$Co}

\def\etal{\hbox{\it\ et al.}}
\def\vol(#1,#2,#3){\ {\bf #1},\ #2\ (#3)}
\def\NP(#1,#2,#3){Nucl.\ Phys.\vol(#1,#2,#3)}
\def\PL(#1,#2,#3){Phys.\ Lett.\vol(#1,#2,#3)}
\def\PLB(#1,#2,#3){Phys.\ Lett.\ B\vol(#1,#2,#3)}
\def\PREP(#1,#2,#3){Phys.\ Rep.\vol(#1,#2,#3)}
\def\PRL(#1,#2,#3){Phys.\ Rev.\ Lett.\vol(#1,#2,#3)}
\def\PR(#1,#2,#3){Phys.\ Rev.\vol(#1,#2,#3)}
\def\ZP(#1,#2,#3){Z. Phys.\vol(#1,#2,#3)}
\def\NIM(#1,#2,#3){Nucl. Instrum. Meth.\vol(#1,#2,#3)}
\def\JLT(#1,#2,#3){J. Low Temp. Phys. \vol(#1,#2,#3)}
\def\ARAA(#1,#2,#3){Ann. Rev. Astron. Astrophys. \vol(#1,#2,#3)}
\def\ARNPS(#1,#2,#3){Ann. Rev. Nucl. Part. Sci. \vol(#1,#2,#3)}

\title{Preliminary Limits on the WIMP-Nucleon Cross Section 
from the Cryogenic Dark Matter Search (CDMS)}

\author{DS~Akerib$^{1,2,}$
\thanks{Presented at TAUP97 Conference, Sept 7-11, 1997, LNGS, Italy.}, 
PD~Barnes~Jr$^3$, 
DA~Bauer$^4$, 
PL~Brink$^5$, 
B~Cabrera$^5$,
DO~Caldwell$^4$, 
RM~Clarke$^5$,
A~Da~Silva$^2$, 
AK~Davies$^5$,
BL~Dougherty$^5$,
KD~Irwin$^5$, 
RJ~Gaitskell$^2$, 
SR~Golwala$^2$,
EE~Haller$^{6,7}$, 
J~Jochum$^2$, 
WB~Knowlton$^{6,7}$, 
V~Kuzminov$^8$,
SW~Nam$^5$, 
V~Novikov$^8$, 
MJ~Penn$^5$, 
TA~Perera$^{1,2}$,
RR~Ross$^{2,6,9}$, 
B~Sadoulet$^{2,6,9}$,
RW~Schnee$^{1,2}$, 
T~Shutt$^2$,
A~Smith$^6$,
AH~Sonnenschein$^4$,
AL~Spadafora$^2$,
WK~Stockwell$^{2,9}$, 
S~Yellin$^4$, 
and BA~Young$^{10}$
\\
{\it $^1$Department of Physics, Case Western Reserve University, 
		Cleveland OH 44106.}
{\it $^2$Center for Particle Astrophysics, 
		University of California, Berkeley CA 94720.}
{\it $^3$Lawrence Livermore National Laboratory, Livermore CA 94550.}
{\it $^4$Department of Physics, University of California, 
		Santa Barbara CA 93106.}
{\it $^5$Department of Physics, Stanford University, Stanford CA 94305.}
{\it $^6$Lawrence Berkeley National Laboratory, Berkeley CA 94720.}
{\it $^7$Department of Material Science and Mineral Engineering,
		University of California, Berkeley CA 94720.}
{\it $^8$Baksan Neutrino Observatory, Institute for Nuclear Research, 
		Russian Academy of Science.}
{\it $^9$Department of Physics, 
		University of California, Berkeley CA 94720.}
{\it $^{10}$Department of Physics, Santa Clara University, 
		Santa Clara CA 95053.}
}%end of author list
       
\begin{document}

\begin{abstract}
We are conducting an experiment to search for WIMPs, or
weakly-interacting massive particles, in the galactic halo using
terrestrial detectors. This generic class of hypothetical particles,
whose properties are similar to those predicted by extensions of the
standard model of particle physics, could comprise the cold component
of non-baryonic dark matter. We describe our experiment, which is
based on cooled germanium and silicon detectors in a shielded
low-background cryostat. The detectors achieve a high degree of
background rejection through the simultaneous measurement of the
energy in phonons and ionization. Using exposures on the order of one
kilogram-day from initial runs of our experiment, we have achieved
(preliminary) upper limits on the WIMP-nucleon cross section that are
comparable to much longer runs of other experiments.
\end{abstract}

\maketitle

\section{Introduction}
%\subsection{Spacing}

Observations of stars and galaxies over a large range of distance
scales indicate the presence of a significant amount of dark matter
that is unseen except for its gravitational effects~\cite{Tr87,KoTu88}.  
There is general
consensus that most of the dark matter, on the order of the critical
density, is comprised of a non-baryonic ``cold'' component.  We are
conducting an experiment to search for direct evidence of WIMPs, or
weakly-interacting massive particles, a generic hypothetical candidate
for cold dark matter.

The experimental challenge is defined in part by considerations of the
early Universe and the properties of our Galaxy. Constraints from the
thermal production of WIMPs in the early universe that yield a
critical WIMP density today are satisfied by particles with masses in
the 10--1000~G\eV\ range and cross sections on the scale of the weak
interaction~\cite{LeWe77}. This range of particle properties suggests
that supersymmetry (SUSY) or other extensions to the standard model
may provide the dark matter~\cite{JKG96}.  If WIMPs exist they would
now make up a major component of the dark matter in our own galactic
halo~\cite{GaTu94}.  For a standard halo comprised of WIMPs with a
Maxwellian velocity distribution characterized by $v_{rms}=270$~km/s
and a mass density of 0.4~GeV/cm$^3$, the expected rate for
WIMP-nuclear scattering is in the range 1--0.001 events per
kilogram of detector per day and the expected recoil energy is as low as 1
keV~\cite{JKG96,SmLe90}.

Despite considerable worldwide efforts,
WIMPs have not yet been detected. Ultimately, experiments have been
dominated by irreducible backgrounds, primarily photons and electrons
from radioactive contamination or activation.
Further progress can be made
by discriminating background events from WIMP events. 
In the CDMS experiment 
rejection of 99\% of the photon background
is achieved 
using detectors that
simultaneously measure the recoil energy in both phonon- and
charge-mediated signals \cite{BCF96,cdmspubs}. 
The ratio of the two measurements
distinguishes electron-recoil events due to background photons from
nuclear-recoil events due to WIMPs since nuclear-recoils are less
ionizing. Following a decade-long development effort, detectors have
now successfully been run in a low-background environment. Our early
data runs yield preliminary upper bounds on the WIMP-nucleon cross
section that are comparable to much longer exposures of other
experiments, illustrating the power of this technique.

\section{Description of the Experiment}

The CDMS detectors employ two distinct technologies for performing the
phonon-mediated measurement of the energy $\Delta E$ deposited in a
scattering event.  One technology uses two
neutron-transmu\-ta\-tion-doped (NTD) germanium thermistors
eutectically bonded to 1.3-cm-thick 6-cm-diameter 165-g cylindrical crystal
of high-purity germanium.  With the device in contact with a 20~mK
bath, monitoring the thermistor resistances gives the temperature rise
$\Delta T = C^{-1} \Delta E$, where $C$ is the heat capacity.
The resulting energy measurement has a FWHM resolution of 
650~eV at 10~keV. The use of two NTDs
permits the rejection of events that originate in one or the other NTDs. 

The other technology uses 
 quasiparticle-trap-assisted 
electrothermal-feedback transition-edge sensors (QETs). 
Tungsten meanders on a surface of a cooled 1-cm-thick cylindrical detector 
are held in the middle of its superconducting transition by
electrothermal feedback using a voltage bias. Deposited energy
drives the tungsten towards normal conduction which produces a current
signal. The time integral of this signal is proportional to the
deposited  energy, which is measured to
650~eV (FWHM) in our 100-gram silicon targets; the technology is now being
transferred to germanium targets. Since the phonon collection time is fast (a
few microseconds), relative-timing 
information from the four sensors on a device
allows a two-dimensional determination of the event position to a few
millimeters. 

The ionization measurement is made by applying a small bias voltage across the
two sides of the semiconductor targets. Electron-hole pairs are
collected efficiently throughout the
bulk of the detectors, but trapping sites near the surface result in a
10--30$\mu$m-thick ``dead layer''  where charge collection is
incomplete.
An energy resolution of 640~eV has been achieved.

The remainder of the apparatus consists of specialized low-activity
detector-housing modules mounted in a cryostat made from a set of
shielded nested copper cans. The cans are cooled by conduction through
a set of concentric horizontal tubes that are connected to a dilution
refrigerator. The cryostat is shielded externally with lead for the
reduction of the gamma rays and polyethylene for the reduction of
neutrons~\cite{Da95}.  Samples of all materials internal to the shield
are carefully screened in a low-background HPGe counting facility for
radio contaminants. Further shielding close to the detectors is
achieved with ancient ultra-low activity lead which has a low
concentration of $^{210}$Pb, a beta-emitter. Due to the complexity of
the detectors and cryostat the first phase of the experiment is being
performed at a shallow site at Stanford University at a depth of 17
meters water-equivalent (mwe). Since cosmic ray muon flux is reduced
by only a factor of 5 at this depth, further rejection of backgrounds
is achieved with a hermetic plastic-scintillator muon veto.

\subsection{Detector Performance}

\begin{figure}
\epsfxsize=7.5cm \epsfbox{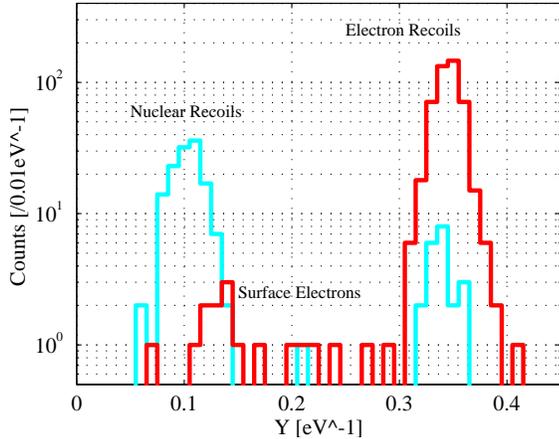} 
\caption[]{\footnotesize {NTD-based germanium detector: charge yield in the recoil-energy range of
15--45~keV for \co\ data (black) and \cf\ data (grey). Gamma
rejection of 99\%is obtained for a nuclear-recoil acceptance of 98\%.}}
\label{e5_cal_hist}
\end{figure}

The capability of the detectors to distinguish photon backgrounds from
WIMP-induced nuclear recoils is demonstrated using photon and neutron
calibration sources; the neutrons serve as test particles to induce
nuclear-recoil events. As discussed above, it is the ratio of the
charge-mediated energy measurement to the phonon-mediated energy measurement
that allows discrimination between electron and nuclear recoils. We
define this ratio as the charge yield $Y$, which is the number of 
electron-hole pairs per eV of recoil energy.  

\begin{figure}
\epsfxsize=7.5cm \epsfbox{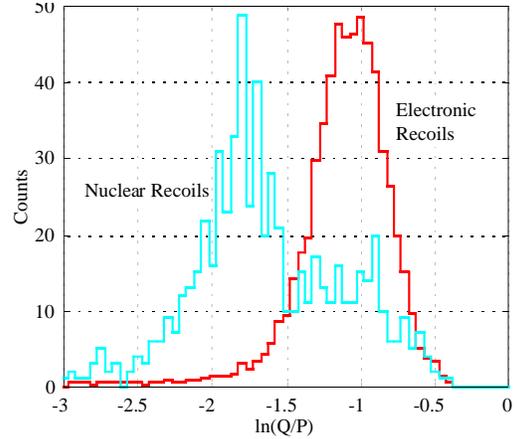}
\caption[]{\footnotesize {QET-based silicon detector: charge yield in the recoil-energy
range of 30--100~keV for \co\ data (black) and \cf\ data (grey).
Gamma rejection of 99\% is obtained for a nuclear-recoil
acceptance of 75\%.}}
\label{f1_cal_hist}
\end{figure}

In separate calibration runs the detectors were alternately exposed to photons
from a \co\ source and neutrons from a \cf\ source. Histograms of the charge
yield are shown in Figures~\ref{e5_cal_hist} and~\ref{f1_cal_hist} for NTD
and  QET detectors, respectively. These data show that 99\% of photon-induced
recoils are rejected 
while high acceptance is maintained for nuclear recoils. 
The events between the main recoil peaks are due to 
electrons that 
deposit energy in
the dead layer and thus have a low charge yield relative to
electron recoils in the bulk.

\section{Low Background Counting}

\subsection{Data Sets}

Several data runs have been taken in the low-background facility over the
past year. We have shown that the experiment can be successfully operated on
month-long timescales with energy resolutions comparable to calibration data.
We have also learned that the rate of photon and neutron backgrounds are
consistent with or less than the expected level, providing important
confirmation that we can reach our goals at the shallow site and that our
screening procedures were effective in limiting these sources of background.

\begin{figure}
\epsfxsize=7.5cm \epsfbox{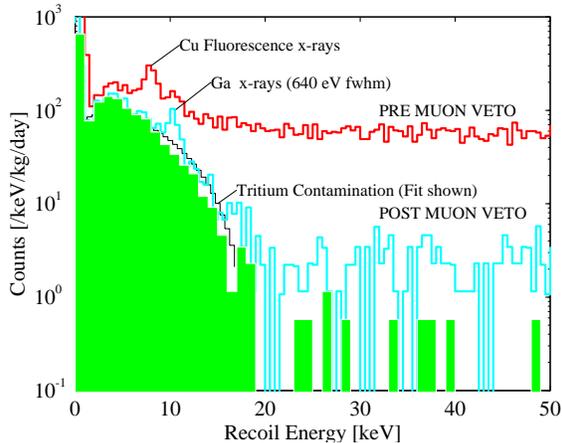}
\caption[]{\footnotesize {Event rate versus recoil energy for a 1.60~kg-d exposure of a 165-g
NTD-based germanium detector. The shaded histogram are events that pass the nuclear recoil cut.}}
\label{b1_r1516_hist}
\end{figure}

The rate versus energy from a 1.60~kg-d exposure of a 165-g NTD-based
germanium detector shows a number of features
(Figure~\ref{b1_r1516_hist}).  The uppermost curve is the full data
set (following event quality cuts) and is dominated by photon events
coincident with the muon veto; the peak at 9~keV is due to florescence
of copper by muon-related photons. The middle curve, events in
anti-coincidence with the muon veto, represents a factor of 20 reduction in
rate. The line at 10~keV is consistent with internal $^{68}$Ge which
undergoes electron capture and leads to a $^{68}$Ga $x$-ray.  The
broad distribution below 18~keV is due electrons from tritium decay on
the surface of the detector. We have also observed a tritium
distribution in events intrinsic to the NTD-Ge thermistors and have
since demonstrated that tritium diffuses out of the NTDs at 550~C,
similar to the temperature used during the eutectic bonding. It should
therefore be possible to control this contamination in future
detectors by baking the NTD prior to bonding.

A cut on charge yield to select nuclear recoils, which is 
based on a fit to neutron calibration data,
results in the
solid histogram in Figure~\ref{b1_r1516_hist}. At low energy, the
spectrum is dominated by the tritium events that have low charge yield
and survive the cut. Above the tritium endpoint the remaining events
are likely due to beta emitters in surface contaminants such as
$^{40}$K from human perspiration or $^{210}$Pb from radon plating.  Further
steps are now being taken to control the contamination by a surface
etch of the detectors late in the fabrication process and through more
careful handling following the final etch (e.g., storage in dry
nitrogen or vacuum). We also expect to reduce our susceptibility to
beta sources external to the detectors by self-shielding them in a
close-packed geometry.  Finally, work is continuing on
eliminating or reducing the dead layer, itself.

\begin{figure}
\epsfxsize=7.5cm \epsfbox{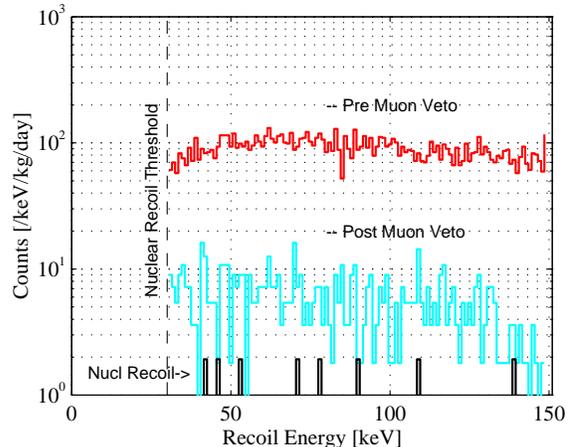}
\caption[]{\footnotesize {Event rate versus recoil energy for a 0.52~kg-d exposure of a 100-g
QET-based silicon detector.}}
\label{f1_r14_hist}
\end{figure}

The energy resolution of the QET-based detector quoted above 
is due to recent 10-fold increase in phonon collection. 
Prior to this improvement, an exposure of
0.52~kg-d was obtained with 
a previous 100-g silicon detector.  Rate versus
energy for these data are 
shown in Figure~\ref{f1_r14_hist}. As with the germanium
data, the muon veto leads to a factor of 20 reduction in rate.
The events in the nuclear-recoil region 
above the threshold of 30~keV are
consistent with the expected number of misidentified gammas. 

\subsection{Preliminary Dark Matter Limits}

We use event rates consistent with nuclear recoils from the two data
sets described above to place an upper limit on the WIMP-nucleon cross
section for spin-independent couplings following reference
\cite{SmLe90}. Figure~\ref{wncs} shows these limits versus WIMP mass
along with other experimental bounds \cite{GeD,NaIR}. Also shown is
the region expected for minimal supersymmetric models (MSSM) that give
a relic density greater than 10\% of the critical density for a Hubble
parameter of 50~km/s/Mpc \cite{JKG96}. Although our exposure is far
less than those of the previous experiments, the sensitivity is
comparable, thus clearly demonstrating the advantage of background
discrimination.

\section{Conclusion and Plans}

Following decade-long development efforts of cryogenic detectors and a
cold shielded low-background environment in which to operate them, we
have begun taking data to search for WIMPs. Data has been taken with two
distinct detector technologies, both of which are on the verge of making major
gains in sensitivity to dark matter. We have identified the primary
background source that presently limits our sensitivity---low-energy electrons
that suffer reduced charge collection.  A number of strategies are in
place
to minimize this and regain the full effectiveness of our event
discrimination technique. Once this has been accomplished, we expect to be
limited by the ambient photon and neutron backgrounds at the Stanford site
with an exposure of about 100-kg-d. To obtain this exposure we will
instrument  two silicon and four germanium devices with QET readout and six
germanium devices with NTD readout, for a total of 200~g of silicon and 2~kg of
germanium.  Comparison of backgrounds in the Ge and Si will provide information on
the backgrounds, especially neutrons. Multiple scattering of neutron backgrounds
in the detector arrays will also provide a handle for background subtraction.

\begin{figure}
\epsfxsize=7.5cm \epsfbox{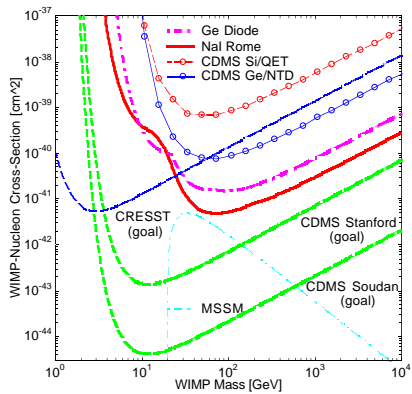}
\caption[]{\footnotesize {The WIMP-nucleon cross section for spin-independent couplings versus
WIMP mass. Upper limits on the cross section are shown for recent runs of CDMS
(preliminary), published results using NaI scintillators and Ge
diodes, and the goals for CDMS and the CRESST experiment. The
dashed region in the lower part of the graph bounds the region where
supersymmetric particles could be the dark matter.}}
\label{wncs}
\end{figure}

In order to take full advantage of these advanced detectors, we plan to continue
the experiment at the Soudan Mine. The 2000 mwe overburden at Soudan will
attenuate cosmic ray muons by some 5 orders of magnitude, which will greatly
reduce cosmogenic activity in the apparatus and greatly reduce the
neutron background. Figure~\ref{wncs} shows the expected sensitivity for a
100-kg-d exposure at the Stanford site and a 5000-kg-d exposure at the Soudan
site. For reference, the projected sensitivity of the CRESST experiment is also
included \cite{cresst}. As seen in the figure, the CDMS experiments will explore a
significant new region of WIMP parameter space, and in particular, a region where
supersymmetric models could provide the dark matter.

\section*{ACKNOWLEDGEMENTS}

This work was supported by the Center for Particle Astrophysics, an NSF
Science and Technology Center operated by the University of
California, Berkeley, under Cooperative Agreement No. AST-91-20005, and
by the Department of Energy under contracts DE-AC03-76SF00098,
DE-FG03-90ER40569, and DE-FG03-91ER40618.  

We gratefully acknowledge the
skillful and dedicated efforts of the technical staffs at LBNL, Stanford
University, UC Berkeley, and UC Santa Barbara.


\begin{thebibliography}{9}
\bibitem{cdmspubs} CDMS papers can be found at
{\tt cfpa. berkeley.edu/group/directdet/gen.html}.
\end{thebibliography}
\end{document}